\documentclass[twocolumn,showpacs,preprintnumbers,amsmath,amssymb]{revtex4}

\usepackage{graphicx}
\usepackage{dcolumn} 
\usepackage{amsmath}
\usepackage{bm}      
\usepackage{graphicx}
\usepackage{color}

\begin{document}

\title {Efficient $6j$ symbol evaluations for atomic calculations}

\author { K.V.P. Latha$^\dagger $, Dilip Angom$^\dagger $, B.P. Das$^\ddagger $}

\affiliation{$^{\dagger}$Physical Research Laboratory, Ahmedabad}
\affiliation{$^{\ddagger}$Indian Institute of Astrophysics,Bangalore}

\begin{abstract}
   We have developed an efficient tabulation scheme to evaluate $6j$ symbols
   for atomic calculations. The scheme is  appropriate for coupled-cluster 
   based calculations. In particular, for perturbed coupled-clusters 
   calculations, which has another perturbation in addition to the residual 
   Coulomb interaction. The scheme relies on the symmetry of the $6j$ symbol
   and the triangular conditions.
\end{abstract}


\maketitle
\section{Introduction}

   The use of angular momentum algebra is inevitable in many body calculations 
of atoms and nuclei. It is essential to couple the single particle states to 
form many particle states or to evaluate matrix elements of tensor operators. 
To couple two angular momentum states, Clebsch-Gordan coefficients are the 
weight factors in the linear combinations of the direct product states. It 
reflects the geometric aspect of the coupled many particle state. Then, for 
three angular momenta,  it is the Wigner $6j$ symbols \cite{wigner} which
are closely related to the Racah recoupling coefficients \cite{racah}.
Higher $3nj$ symbols are required to couple larger number 
of angular momentum states \cite{varsha}. As mentioned earlier, these 
coefficients also occur while evaluating the matrix elements of tensor 
operators in the angular momentum basis. In atomic many body theory, $6j$ 
symbols occur frequently while evaluating the matrix elements of the 
two-electron Coulomb interaction \cite{lindgren}. The number of times $6j$s 
symbols are calculated increases substantially in structure or properties 
calculations, where the the Coulomb interaction is treated perturbatively 
to high orders or non-perturbatively to all orders.  

  In this paper, we address the $6j$ symbol evaluation requirements of 
coupled-cluster calculations, a non--perturbative many body theory 
\cite{bartlett}, of heavy atoms. The theory is applicable to both, atomic 
structure and properties calculations. In these calculations, depending on the 
number of orbitals in the basis set chosen, the number of $6j$ symbol
 evaluations could be as large as $10^9$. It is found that, the $6j$ symbol
evaluations of this magnitude take a large fraction, in many instances about 
30--40 \%, of the total computational time. This has severe implications in 
coupled-cluster calculations, where the working equations are a set of 
non linear algebraic equations. The number of $6j$ symbol evaluations increases
manifold in perturbed coupled-cluster theory, where the perturbation
is an operator of rank one or higher in the electron space. The theory is 
appropriate to calculate the effects of discrete symmetry violations in atoms
\cite{sahoo,latha-08} and transition properties. The large number of $6j$ 
symbol in the perturbed cluster equations, set of linear algebraic 
equations, is a serious issue for calculating the properties of heavy atoms. 
Further, the evaluations are repetitive as iterative schemes suitable method 
to solve the equations.

  One solution to avoid the repeated evaluations of the $6j$ sybmols and 
reduce the time of calculation is, evaluate all the needed $6j$ symbols and 
tabulate it. However, this is easier said than done. The possible number of 
$6j$ symbols upto a maximum value of angular momentum  $j_{\rm max}$ grows 
rapidly as $j_{\rm max}$ is increased. Then the tabulation require large 
arrays, which is an undesirable feature in large scale computations. Employing 
the symmetry properties, it is possible to reduce the number significantly. But 
the disadvantage of incorporating the symmetries is, large number of binary 
operations are essential to retrieve the tabulated values. An optimal 
scheme is to tabulate with selected symmetry properties. The other approach
is to improve efficiency of the $6j$ symbol calculations is to employ a fast
evaluation scheme \cite{wei-99,wei-04}. This scheme would be faster than
the calculations with factorials. But it involves several binary 
operations as it has a summation, so the tabulation is a better choice.

 The paper is organized as follows in Section.\ref{rec_coeff}, the expression
and properties of $6j$ symbols are described in brief. Then in 
Section.\ref{6j_cc} the reason for larger number of $6j$ symbols in 
perturbed coupled-cluster calculations is discussed with an example. We 
have chosen an example from electric dipole moment calculations. Then 
the method of tabulation and retrieval we have developed are given in 
Section.\ref{stor_retr}. This is followed by results and discussions, and
conclusions.


\section{6j symbols}
  \label{rec_coeff}

\subsection{Symmetry and representation}

  Consider a many particle system consisting of three particles, each having 
individual angular momenta $j_1$, $j_2$ and $j_3$. Then the total
angular momentum $J$ of the system is the vector sum of $j_1$, $j_2$ and 
$j_3$. 

  In general, there are three distinct ways of coupling the three 
angular momenta. These are $|J_{12},j_3,JM\rangle $, $|J_{13},j_2,JM\rangle $
and $|j1,J_{23},JM\rangle $, where $J_{ij}$ represents coupling of 
$j_i$ and $j_j$, and $JM$ is the total angular momentum. Among these 
possibilities, states of two different coupling schemes are related through 
a unitary transformation
\begin{equation}
  |j_1 J_{23},JM\rangle = \sum_{J_{12}} |J_{12},j_3,JM\rangle 
 \langle J_{12},j_3,JM |j_1, J_{23},JM\rangle .
\end{equation}
The elements of the transformation are the recoupling coefficients 
$\langle\cdots|\cdots\rangle$. The symmetric representation of these
coefficients is 
\begin{eqnarray}
    \langle J_{12},j_3,JM |j_1, J_{23},JM\rangle &= &
   (-1)^{j_1+j_2+j_3+J} \left[J_{12},J_{23}\right] \times \nonumber \\
     && \left\{\begin{array}{ccc} 
          j_1 & J_{12} & j_2 \\ 
          j_3 & J_{23} & J \end{array}\right\},
\end{eqnarray}
where $\{\ldots\}$ is the $6j$ symbol. It is invariant with respect to any 
column permutation and rows interchange for a pair of columns, which 
explains why it is the symmetric representation of recoupling coefficients.
Counting the number of invariant transformations, column permutations and 
row interchange, one $6j$ symbol has 24 equivalent representations. Then, the 
angular momenta in a $6j$ symbol satisfy the triangular conditions
\begin{subequations}
\begin{eqnarray}
   |j_1 - j_2| & \leqslant J_{12}  \leqslant & j_1 + j_2,  \\
   |j_2 - j_3| & \leqslant J_{23}  \leqslant & j_2 + j_3,  \\
   |J - j_3|   & \leqslant J_{12}  \leqslant & J + j_3,   \\
   |J - j_1|   & \leqslant J_{12}  \leqslant & J + j_1 .  
\end{eqnarray}
 \label{6j_ineq}
\end{subequations}
The $6j$s symbols are products of four Clebsch-Gordan coefficients or 
equivalently four $3j$ symbols, the symmetric representation of Clebsch-Gordan 
coefficients. Besides coupling of angular momenta, the $6j$ symbols are part 
of the angular factors in the matrix elements of coupled tensor operators. 

\begin{figure}[h]
   \includegraphics[width=3cm]{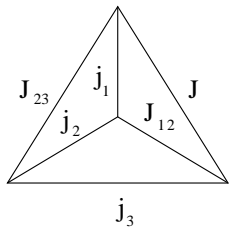}
   \caption{Diagrammatical representation of $6j$ symbol.}
   \label{6jfig}
\end{figure}
  Atomic structure and properties calculations involve calculations of 
matrix elements of a tensor operator in the single particle basis, which
comprises of the radial and angular parts. The evaluation of the angular 
part involves the expansion of 
the matrix elements using the Wigner-Eckart theorem and 
rearranging the angular momenta couplings and tensor coupling of the 
operators. It then simplifies to product of $3j$ and $6j$ symbols,
phase factors and constant factors. This procedure is sometimes tedious 
especially when the matrix element has several angular momenta and 
operators. A simpler way is to use angular momentum diagrams. In atomic 
many-body theory, these are the Goldstone diagrams where operator and angular
momenta representations replace the interaction and orbital lines respectively
\cite{lindgren}. The diagrams are then simplified and evaluated based on
rules which separate close or open parts to diagrammatic representation of
angular momenta coupling identities. In this scheme, the diagrammatic 
representation of $6j$ symbols is shown in Fig. \ref{6jfig}  
The diagram has a high degree of symmetry, which are equivalent to the
symmetry of the algebraic representation. It is non-zero only if the 
angular momenta meeting at each vertex satisfy the triangular conditions
listed in the earlier section.


\subsection{$6j$ symbol calculation and unique representation}

  The $6j$ symbols are  normally calculated using the Racah formula
\begin{eqnarray}
\left\{\begin{array}{lll} 
          j_1 & j_2 & j_3 \\ 
          l_1 & l_2 & l_3 
       \end{array}\right\}
   &= & \sqrt{\Delta(j_1j_2j_3)\Delta(j_1l_2l_3)\Delta(l_1j_2l_3)} \times
           \nonumber \\
     && \sqrt{\Delta(l_1l_2j_3)}\sum_t\frac{(-1)^t(t+1)!}{f(t)},	
 \label{6jformula}
\end{eqnarray}
where $\Delta(abc)$ is a triangle coefficient
\begin{equation}
   \Delta(a, b, c) = \frac{(a+b-c)!(a-b+c)!(-a+b+c)}{(a+b+c+1)!}
\end{equation}
and 
\begin{eqnarray}
 f(t)&=& (t-j_1-j_2-j_3)!(t-j_1-l_2-l_3)!  \nonumber \\
     &&  \times (t-l_1-j_2-l_3)! (t-l_1-l_2-j_3)!  \nonumber \\ 
     &&  \times (j_1+j_2+l_1+l_2-t)! (j_2+j_3+l_2+l_3-t)! \nonumber \\
     &&  \times (j_3+j_1+l_3+l_1-t)!.
\end{eqnarray}
The summation is over all integer values of $t$ for which $f(t)$ is defined.
In other words, values of $t$ which make the arguments of factorials 
in $f(t)$ non negative. Often, theoretical atomic and nuclear physics 
calculations require several values of $6j$ symbols, for which one can refer to 
one of the several  published tabulations. Usually, in these tables, 
as mentioned earlier each $6j$ symbol has twenty four equivalent 
representations, only one of the representations is listed. 
\begin{equation}
   \left\{\begin{array}{lll}
             j_1 \rightarrow & j_2 \rightarrow & j_3      \\
             \downarrow      & \downarrow      & \!\!\!\!\!\!\!\!\!\searrow \\
             l_1             & l_2             & l_3    
          \end{array}\right\}
  \label{unique_6j}
\end{equation}
A unique choice of selecting
the one representation is to impose the inequalities $j_1 \geqslant j_2$, 
$j_2 \geqslant j_3$, $j_1 \geqslant l_1$, $j_2 \geqslant l_2$ and 
$j_2 \geqslant l_3$. The relations  are symbolically represented in 
Eq.(\ref{unique_6j}), where $\cdots\rightarrow\cdots $ denotes 
$\cdots \geqslant \cdots$.


\section{$6j$ symbols in Coupled-cluster calculations}
  \label{6j_cc}

\begin{figure}[h]
   \includegraphics[width=6cm]{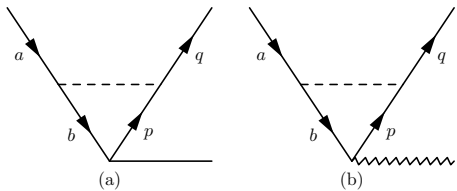}
   \caption{One of the single excitation diagram, which involves contraction
            of single excitation cluster operator and $V_{\rm es}$, the 
            residual Coulomb interaction. The solid and zigzag
            lines represent cluster operators arising from $V_{\rm es}$ and 
            $H_{\rm PTV}$ respectively. In diagram (a) the cluster operator
            zero rank. Whereas in (b) has rank one.}
   \label{ccedm-ex}
\end{figure}

Coupled-cluster theory \cite{bartlett} is considered to be one of the most
accurate many-body theory. This is evident from the fact that it is an
all order theory and proved through extensive calculations in atom, molecules
and nuclei. Recently, we have developed a coupled-cluster based method to 
calculate electric dipole moments of closed-shell atoms \cite{latha-08}. The 
theory has cluster amplitudes arising from two interaction Hamiltonians. First,
the residual Coulomb interaction $V_{\rm es}$ and second, the discrete 
symmetry violating interaction $H_{\rm PTV}$. As a result, the angular parts 
of the cluster equations have large number of $6j$ symbols. To demonstrate, 
consider the single excitation diagram with $V_{\rm es}$ as interaction shown 
in Fig.\ref{ccedm-ex}. The equivalent algebraic expression is
$$
  \langle bp|\frac{1}{r_{12}}|aq\rangle\times \langle q|O_1^{(1)}|b \rangle
$$
where $O^{(1)}$ is a rank zero and one cluster operators for $V_{\rm es}$ and
$H_{\rm PTV}$ respectively, $a$ and $b$ are the occupied orbitals, and 
$p$ and $q$, are the virtual orbitals.
The angular factor of the diagram (a) in the the figure is 
\begin{equation}
      (-1)^{j_b-j_a+k}\sqrt{2j_b + 1}\delta(j_a, j_q)\delta(j_b, j_p),    
  \label{cct0_ang}
\end{equation}
where $k$ are the allowed multipoles of the $V_{\rm es}$ interaction and $j_i$
are the angular momenta of the orbitals. This angular factor, consisting of
phase factor, a constant and Kronecker delta is computationally not
demanding.  However, for the diagram (b), where the cluster operator is
rank one, the angular factor is
\begin{equation}
     (-1)^{j_a + j_p + 1} \left \{ \begin{array}{ccc}
                                      j_b & j_q & 1 \\
                                      j_p & j_a & k
                                   \end{array} \right \}.
  \label{ccedm_ang}
\end{equation}
From the expression of $6j$ symbol, given in Eq.(\ref{6jformula}), it's 
evaluation has far larger number of arithmetic operations than calculation of 
Eq.(\ref{cct0_ang}). The present comparison is for one of the simpler diagrams.
The occurrence of $6j$ symbol is larger in diagrams of doubles and more 
complicated. Calculating a larger number of $6j$ symbols to solve the perturbed
cluster amplitude equations is a serious performance issue. It is particularly 
severe for properties calculations of heavy atoms, the number of cluster 
amplitudes is in millions. Another factor adding to the inefficiency is the 
repeated occurrence of the same $6j$ symbol in several diagrams, the multiple 
evaluations is expensive. A trivial solution is tabulating the $6j$ symbols,
however this is not simple to implement.


\section{Storage and retrieval}
   \label{stor_retr}

\subsection{Symmetry considerations}

   The optimal basis sets chosen for accurate structure and properties 
calculations of heavy atoms have single particle wave functions of high 
angular momenta. For example, for structure and properties calculations 
of atomic ytterbium, in $jj$ coupled scheme.  The optimal basis set
consist of  orbitals upto $h$ symmetry, which has angular momenta $9/2$ and 
$11/2$. This is an important consideration to describe the 
electron-electron correlation energies accurately. An immediate outcome is
the large values of the angular momenta in the angular part of the 
matrix elements. Then, the number of possible $6j$ symbols which can occur is 
extremely large. For a basis set consisting of orbitals upto $h$ symmetry,
the maximum angular momenta which can occur in $6j$ is $11$. That is the 
maximum rank of the operator which satisfies triangular condition for 
matrix elements between two $h$ orbitals. The approximate total number of 
$6j$s is then $22^6\approx 1.2\times 10^8$, which is a large number.

   One option to reduce the number of entries in the tabulations is to
impose the inequalities, diagrammatically shown in Fig.\ref{unique_6j}, 
to the angular momenta of the $6j$ symbol. Such a scheme reduces the number by 
a factor of 24, the number of equivalent forms.  Another option is to apply 
the four triangular  conditions in Eq.(\ref{6j_ineq}). However, as discussed
in the results sections, the exact implementation of these conditions is 
computationally inefficient. An optimal selection of the inequalities
from Eq.\ref{unique_6j} are: $j_1\geqslant l_1$, $j_1\geqslant j_2$,
$j_1\geqslant j_3$ and $j_2\geqslant l_2$.  These reduces the
number of equivalent forms reasonably.


\subsection{Tagging}

  Tabulation of the $6j$ symbol during computational calculations imply tagging
each one with a unique integer and a scheme to evaluate the tag 
efficiently. The other considerations are, the tags be in a sequence 
and preserve the inequality conditions. A straight forward scheme is 
to take advantage of the three inequalities $j_1\geqslant l_1$,
$j_1\geqslant j_2$ and $j_2\geqslant l_2$. Tagging each
$(j_1, l_1)$ pair is equivalent to indexing the elements of the lower 
triangular matrix and the integer tag is $2j_1(2j_1 + 1)/2 + 2l_1$. The 
multiplication of $j_1$ and $l_1$ by two is essential as the angular momenta 
are in multiples of half. Similarly, the $(j_2, l_2)$ are tagged. Further,
extending the scheme, the pairs $(j_1, l_1)$ and $(j_2, l_2)$ is mapped to 
a single integer number, which is like a super tag of $(j_1, j_2, l_1, l_2)$
combinations. For each tag, the possible $(j_3, l_3)$ are considered. 
Structurally, this can visualized as a stack of matrices $(j_3, l_3)$,
one for each $(j_1, j_2, l_1, l_2)$.
\begin{figure}[h]
   \includegraphics[width=7cm]{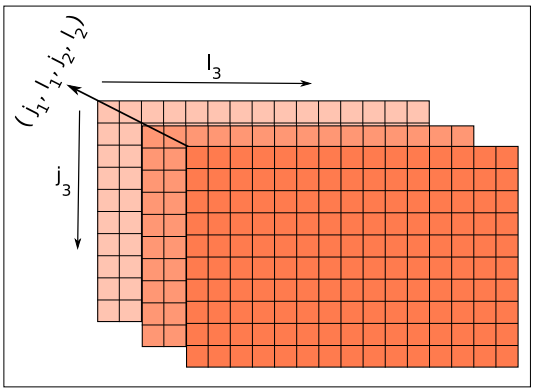}
   \caption{The $(j_3, l_3)$ pairs for a specific $(j_1, l_1, j_2, l_2)$
            combination is like a stack, each slot in one of the stacks
            represents one unique $(j_1, l_1, j_2, l_2, j_3, l_3)$ 
            combination or $6j$.
            }
   \label{stack}
\end{figure}
An important point is, unlike $(j_1, l_1)$ and $(j_2, l_2)$ pairings,
tagging $(j_3, l_3)$ pairs is equivalent to indexing a full matrix. This
follows from the absence of inequality between $j_3$ and $l_3$. From 
the inequalities adopted for tagging, the maximum number of $(j_3, l_3)$ 
pairs is $(2j_1+1)(2j_1+1)$. When the $6j$ are stored in a one dimensional
array, to retrieve one $6j$, skip the locations of previous 
$(j_1, l_1, j_2, l_1)$ combinations and then from $(j_3, l_3)$ evaluate  
the offset within the stack.


\subsection{Efficiency}

  A rough estimate of the efficiency of the tabulation and retrieval, 
compared to the actual calculation is to compare the number of binary 
operations in the two schemes. This include arithmetic and boolean binary
operations. The number of binary operations to calculate $6j$ symbol from 
is estimated from Eq. \ref{6jformula}. The number of binary operations 
required to calculate each $\Delta(abc)$ is thirteen. In total, to 
evaluate the product of the four  $\Delta(abc)$ require fifty six binary 
operations arithmetic operations. Then the, for one $t$, the number 
calculation of  $f(t)$ require twenty four binary operations. Considering
the summation over $t$, the total number of binary operations is 
$\approx 56 + 28 \times (\sigma +1)$, where $\sigma$ is the number of 
values $t$ can have. This is without the binary operations to calculate
the factorials,  the assumption is these are precalculated.
In comparison, for the scheme outlined to store and retrieve $6j$ symbols, the 
number of binary operations required to retrieve the $6j$ symbols from memory 
is $\approx $ 50. This involves the arrangement of the six $j$ symbols to 
satisfy the inequality conditions adopted and calculation of the tag. 
Approximately, the tabulation and retrieval of $6j$ symbol is like calculating
without the evaluation of $f(t)$. This indicates a significant gain in 
performance.

 
\section{Results and discussions}

\begin{figure}[h]
   \includegraphics[width=9cm]{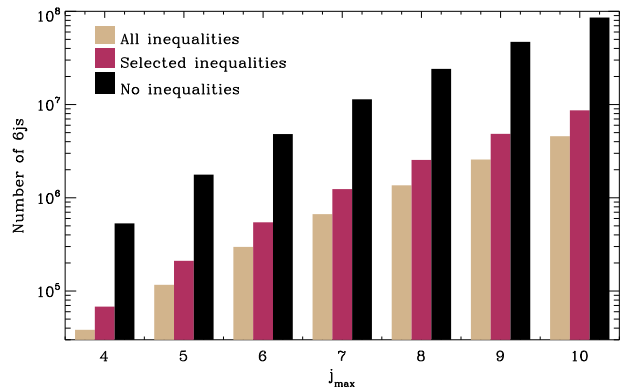}
   \caption{Number of possible angular momenta combinations to form a $6j$ 
            symbol with a maximum angular momentum $j_{\rm max}$. Black, maroon
            and tan are the number of $6j$ symbols when no inequality, selected 
            inequality and all inequalities are imposed to the angular 
            momenta.}
   \label{6j_num}
\end{figure}

\begin{figure}[h]
   \includegraphics[width=9cm]{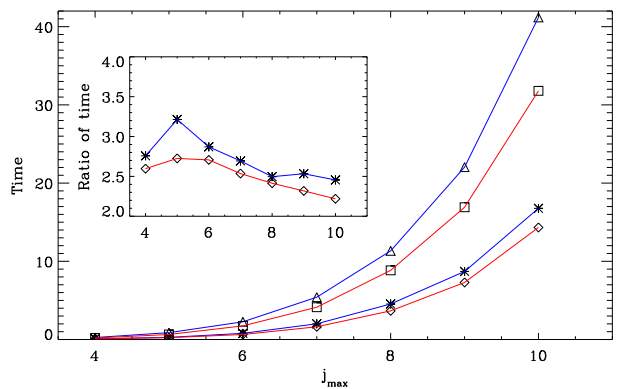}
   \caption{Run time to calculate $6j$ symbols with all the possible 
            combinations upto the maximum angular momenta $j_{\rm max}$. The 
            pair of curves at the bottom, connecting the $*$ and $\diamond$ 
            correspond to retrieval from the tabulated list. The pair at 
            the top, $\vartriangle$ and $\Box$ correspond to numerical 
            calculation from the Racah expression. The red and blue curves 
            are with and without 
            optimizations. The inset curves shows the ratio of the two.
           }
   \label{perf_notria}
\end{figure}

   To quantify the relative computational efficiency, between calculating the 
$6j$ symbols from Racah formula Eq.(\ref{6jformula}) and retrieving 
from a tabulated list, we compare the execution time of the two schemes
for calculating a set of $6j$ symbols. The set chosen consists of $6j$ symbols
with all possible combinations of angular momenta upto a maximum angular 
momentum $j_{\rm max}$. The number of the $6j$ symbols with and without the 
imposition of the symmetry conditions are shown in Fig.\ref{6j_num}.
The plots in the Fig.\ref{perf_notria} shows the plots of computation time for 
the values of $j_{\rm max}$ ranging from $4$ to $10$. 
The inset plots in Fig.\ref{perf_notria} is the ratio of the computation time
to calculate the $6j$ symbols from Racah formula and retrieve from the 
tabulated list. The ratio has a maximum of $\sim$3.4 around $j_{\rm max}=5$ and 
is 2.5 for the largest $j_{\rm max}=10$ case. There is a significant 
performance gain. For the largest $j_{\rm max}$, which is relevant for
the coupled cluster calculations of heavy atoms, there is performance gain of
$\sim$220\% gain after compiler optimizations.
Performance of the tabulation is enhanced further when an approximate form
of the triangular conditions of the $6j$ symbols in Eq.(\ref{6j_ineq}) are 
imposed.  The approximate form is to ensure that the sum of the three angular 
momenta, which should satisfy triangular condition, is integer. For example, 
consider the first inequality 
$ |j_1 - j_2| \leqslant J_{12}  \leqslant  j_1 + j_2 $. Instead of imposing
the inequality, we check if the sum $j_1 + j_2 + J_{12}$ is integer. The
retrieval scheme proceeds if it is integer, otherwise it returns a zero value.
This improves the efficiency of the retrieval scheme by avoiding the steps to 
transform the arguments.
\begin{figure}[h]
   \includegraphics[width=9cm]{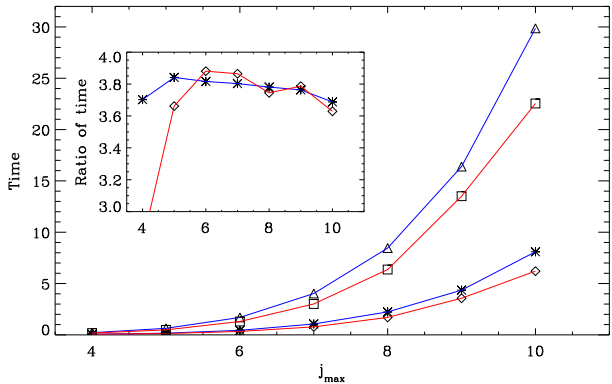}
   \caption{Run time to calculate $6j$ symbols with all the possible 
            combinations upto the maximum angular momenta $j_{\rm max}$ and 
            imposing the triangular conditions approximately. The lower pair of 
            curves connecting the $*$ and $\diamond$ correspond to retrieval
            from the tabulated list. The upper pair, $\vartriangle$ and 
            $\Box$ correspond to numerical calculation from the 
            Racah expression. The red and blue curves are with and without 
            optimizations. The inset curves shows the ratio of the two.
            }
   \label{perf}
\end{figure}
A comparison of the run time of calculating $6j$ symbols and retireving from 
the tabulated values is shown in Fig.\ref{perf}. There is a marked improvement,
retrieving from the table is more than 350\% faster than the calculation. 
This is evident from the inset plot in Fig.\ref{perf}. In coupled-cluster
calculations of heavy atoms, the number of times $6j$ symbol is 
evaluated is extremely large. For example, consider the cluster amplitude 
calculations of Hg with all the core orbitals and one virtual orbitals from 
each symmetry. At the linear level this calculation require 
$\approx 2.9 \times 10^8$ evaluations of $6j$ symbol. The number is higher for 
full scale coupled-cluster calculations and  much higher in perturbed
coupled-cluster calculations $\approx 7.5\times 10^9$. In the later case, 
perturbed coupled-cluster, implementing the tabulation scheme provides a 
$\approx 30$\% performance improvement. In other words, calculation of $6j$
symbols takes $\approx 30$\% of the total run time of coupled-cluster 
calculation, with tabulation this is reduced to less than 1\%.

 
\section{Conclusions}

  Tabulating the $6j$ symbols coefficients improves the efficiency of the 
coupled-cluster computations significantly. The reduction of binary operations 
in the tabulation scheme, as explained earlier, accounts for the improvement.
The optimal scheme of tabulation and retrieval is to impose a restricted
set of symmetry properties and approximate triangular condition. Strict 
implementation of the symmetry properties and triangular conditions 
compromises the efficiency of the tabulation scheme. In terms of evaluating
the coefficients, the tabulation scheme is more than 350\%  faster than
the actual evaluation. This translates to reducing more than 30\% run time 
of coupled-cluster calculations.


\begin{acknowledgments}
The work was supported by the DST-NSF project No. DST/INT/US-NSF/RPO-154/04. 
One of the author, DA would like to thank the Dmitry Budker from University
of Califorrnia, Berkeley for the hospitality and stimulating discussions
during his recent visit.
\end{acknowledgments}

\end{document}